\documentclass[12pt]{article}

\usepackage{amsmath,amssymb}

\newcommand{\be}{\begin{equation}}
\newcommand{\ee}{\end{equation}}

\renewcommand{\theequation}{\arabic{section}.\arabic{equation}}

\def\Tr{{\rm Tr}\,}
\def\cN{{\cal N}}

\def\bea{\begin{eqnarray}}
\def\eea{\end{eqnarray}}
\def\nn{\nonumber}

\def\theequation{\arabic{section}.\arabic{equation}}

\begin{document}
\begin{titlepage}
\begin{center}
{\Large\bf
Effective action of three-dimensional \\[1mm]
extended supersymmetric matter \\[2.5mm]
on gauge superfield background}

\vspace{1cm}

{\large\bf I.L. Buchbinder$\,{}^{\dag}$,
N.G. Pletnev$\,{}^{\ddag}$,
I.B. Samsonov$\,{}^{\star}$
\\[8pt]
\it\small $^\dag$Department of Theoretical Physics, Tomsk State
Pedagogical University,\\ 634061 Tomsk, Russia, {\tt email:\ joseph@tspu.edu.ru}
\\[8pt]
$^\ddag$Department of Theoretical Physics, Institute of Mathematics,
630090 Novosibirsk, Russia\\
{\tt email:\ pletnev@math.nsc.ru}\\[8pt]
$^\star$Laboratory of Mathematical Physics, Tomsk Polytechnic University,
634050 Tomsk, Russia\\
{\tt email:\ samsonov@mph.phtd.tpu.ru}}
\end{center}
\vspace{0.5cm}

\begin{abstract}
We study the low-energy effective actions for gauge superfields
induced by quantum $\cN=2$ and $\cN=4$ supersymmetric matter fields
in three-dimensional Minkowski space.
Analyzing the superconformal invariants in the $\cN=2$
superspace we propose a general form of the $\cN=2$ gauge invariant and superconformal
effective action. The leading terms in this action are fixed by the symmetry
up to the coefficients while the higher order terms with respect
to the Maxwell field strength are found up to one arbitrary
function of quasi-primary $\cN=2$ superfields constructed from the
superfield strength and its covariant spinor derivatives. Then we
find this function and the coefficients by direct quantum
computations in the $\cN=2$ superspace. The effective action of
$\cN=4$ gauge multiplet is obtained by
generalizing the $\cN=2$ effective action.
\end{abstract}

\end{titlepage}

\setcounter{equation}{0}
\section{Introduction}
Modern interest to three-dimensional supergauge models with extended
supersymmetry is motivated mainly by recent progress in constructing
and studying the field theories describing the worldvolume degrees
of freedom of M2 branes. Such models are usually referred to as the
Bagger-Lambert-Gustavsson (BLG) \cite{BLG} and
Aharony-Bergman-Jafferis-Maldacena (ABJM) \cite{ABJM} theories
which are the superconformal Chern-Simons-matter models with $\cN=8$ and
$\cN=6$ supersymmetry, respectively. Since the superconformal
symmetry is preserved on the quantum level, these theories are
dual to the superstring theory on the corresponding background
within the AdS$_4$/CFT$_3$ correspondences.

One of the general
problems for the BLG and ABJM models is to study the effective action
which would describe an effective quantum dynamics of M2 branes.
In particular, such effective actions receive contributions in the
gauge field sector induced by quantum matter fields
which can be studied independently of the other contributions.
A good starting point for understanding this general issue is
the effective action for the Abelian gauge superfield induced by
quantum matter superfields.

In the present paper we explore the three-dimensional
supersymmetric
Euler-Hei\-sen\-berg-type effective action which appears as a result
of one-loop contributions from quantum supersymmetric matter.
This problem is interesting
not only from the point of view of BLG and ABJM models, but also as
a part of the effective action in the three-dimensional
supersymmetric electrodynamics. In the non-supersymmetric case
this problem was studied in \cite{Redlich}, but superspace
analysis has never been done (c.f.\ the effective action in the
four-dimensional supersymmetric
electrodynamics which was studied in superspace in
\cite{d4eff,Kuz03,Kuz07}). In the present paper we fill this gap
by deriving the Euler-Heisenberg effective actions for the
model of $\cN=2$ chiral superfield and $\cN=4$ charged hypermultiplet
interacting with the background gauge superfields.

In our work we employ the $\cN=2$, $d=3$ superspace
approach which is similar to the $\cN=1$, $d=4$ superspace.
In particular, the $\cN=2$, $d=3$ chiral and vector multiplets appear by
dimensional reduction from the four-dimensional $\cN=1$
supersymmetric ones while the hypermultiplet and the $\cN=4$
vector multiplet in three dimensional Minkowski space
originate from the $\cN=2$, $d=4$ hypermultiplet
and gauge superfield, respectively.
We consider the background $\cN=2$, $d=3$ gauge superfield constrained
by $D_\alpha W_\beta=D_{(\alpha} W_{\beta)}=\rm const$, where
$W_\alpha$ is the superfield strength. In components, this
constraint corresponds to the constant Maxwell field strength,
$F_{mn}=\rm const$.

As soon as the classical action of the chiral
superfield in the background gauge superfield is superconformal, the
resulting effective action should be superconformal as well. We show
that the gauge and superconformal invariance restrict the functional
form of the leading terms in the effective action
uniquely, up to coefficients, while the higher order terms with
respect to the Maxwell field strength are encoded in a single
arbitrary function of one superconformal quasi-primary superfield.
Then we find this function as well as the coefficients by
direct quantum computations in the $\cN=2$, $d=3$ superspace. A
straightforward generalization of these results to the $\cN=4$
case leads to the effective action of
the $\cN=4$ charged hypermultiplet interacting with background
gauge superfield.

The paper is organized as follows. We begin Section 2 with a
short review of the chiral superfield model in the $\cN=2$ superspace
and specify the constraints on the background gauge superfield
under considerations. Then we discuss general structure of the
gauge-superfield-dependent $\cN=2$ supersymmetric effective action
subject to the constraints of gauge and superconformal invariance.
In Section 3 we compute the one-loop effective
actions in the models of $\cN=2$ chiral superfield interacting with
the background gauge superfield
as well as for the $\cN=4$ charged hypermultiplet
using the Fock-Schwinger's proper-time technique
in the $\cN=2$ superspace. In the last section we discuss the obtained
results and their possible generalizations. Appendix A
contains basic formulae concerning the $\cN=2$, $d=3$ superspace in
our conventions. In Appendix B we consider a representation of the
superconformal group on the superfields in the $\cN=2$ superspace.

\setcounter{equation}{0}
\section{General structure of superconformal effective action
in $\cN=2$ superspace}
\subsection{Classical action of chiral superfield interacting
with the gauge superfield}
In this subsection we review some features of the $\cN=2$, $d=3$
chiral and gauge superfield models which will be used in
the next sections. Our conventions for the $\cN=2$ superspace are
collected in the Appendix A.

Let us consider a classical action for the
chiral superfield $Q$ interacting with
the Abelian background gauge superfield $V$,
\be
S_{\cN=2}=-\int d^3x d^4\theta \, \bar Q e^{2V}Q\,,
\label{S}
\ee
which is invariant under the following gauge transformations
\be
Q\to e^{i\Lambda} Q\,,\quad
\bar Q \to \bar Q e^{-i\bar\Lambda}\,,\quad
e^{2V}\to e^{i\bar\Lambda} e^{2V} e^{-i\Lambda}\,,
\label{gauge-symm}
\ee
with $\Lambda$ and $\bar\Lambda$ being (anti)chiral superfield gauge
parameters.
The chiral multiplet consists of the complex scalar $f$,
complex spinor $\psi_\alpha$ and complex auxiliary scalar $F$,
\be
Q=f+\theta^\alpha\psi_\alpha+\theta^2 F
+i\theta^\alpha\bar\theta^\beta \partial_{\alpha\beta}f
+\frac i2\theta^2\bar\theta^\alpha\partial_{\alpha\beta}\psi^\beta
+\frac14\theta^2\bar\theta^2\square f\,.
\ee
The vector multiplet in three dimensions is built from one real
scalar $\phi$, one complex spinor $\lambda_\alpha$, one vector field
$A_{\alpha\beta}=\gamma^m_{\alpha\beta}A_m$ and one real auxiliary
scalar $D$. In the Wess-Zumino gauge the component decomposition
for $V$ is given by
\be
V=\theta^\alpha\bar\theta^\beta A_{\alpha\beta}
+ i\theta^\alpha\bar\theta_\alpha\phi
+ i\theta^2\bar\theta^\alpha\bar\lambda_\alpha
- i\bar\theta^2\theta^\alpha\lambda_\alpha
+\theta^2\bar\theta^2 D\,.
\label{V}
\ee

It is important to specify the background gauge superfield under
considerations. In general, the vector multiplet arises within
standard geometric approach based on covariantization of
the flat superspace derivatives,
\be
D_\alpha\to\nabla_\alpha=D_\alpha+{\bf A}_\alpha\,,\quad
\bar D_\alpha\to\bar\nabla_\alpha=\bar D_\alpha+\bar {\bf A}_\alpha\,,\quad
\partial_m\to\nabla_m=\partial_m+{\bf A}_m\,,
\label{covD}
\ee
where the following superfield constraints are imposed \cite{HKLR,ZP,NG}
\bea
\{\nabla_\alpha,\bar\nabla_\beta \}&=&-2i(\gamma^m)_{\alpha\beta}
\nabla_m +2i\varepsilon_{\alpha\beta}G\,,
\label{alg1}
\\
{}[\nabla_\alpha,\nabla_m]&=&-(\gamma_m)_{\alpha\beta}\bar
W^\beta\,,\qquad
[\bar\nabla_\alpha,\nabla_m]=(\gamma_m)_{\alpha\beta}
W^\beta\,,
\label{alg2}
\\
{}[\nabla_m,\nabla_n]&=&i{\bf F}_{mn}\,.
\label{algebra}
\eea
The superfield strengths in the rhs in
(\ref{alg1})--(\ref{algebra}) satisfy the following reality
properties
\be
G^*=G\,,\quad
(W^\alpha)^*=\bar W^\alpha\,,\quad
({\bf F}_{mn})^*={\bf F}_{mn}\,.
\ee
As usual, there are many Bianchi identities for these
superfield strengths which are important for studies of the
effective action and quantization. In particular, the superfield
strengths $W_\alpha$ and $\bar W_\alpha$ are (anti)chiral,
\be
\bar D_\alpha W_\beta=0\,,\qquad
D_\alpha \bar W_\beta=0\,,
\ee
and obey
\be
D^\alpha W_\alpha=\bar D^\alpha \bar W_\alpha\,.
\ee
An important feature of the $\cN=2$, $d=3$ superspace formulation
of the gauge multiplet is that the superfield strengths
$W_\alpha$, $\bar W_\alpha$ are expressed in terms of the scalar
superfield strength $G$,
\be
W_\alpha=\bar D_\alpha G\,,\qquad
\bar W_\alpha= D_\alpha G\,,
\label{W(U)}
\ee
subject to the following constraints
\be
D^2 G=0\,,\qquad \bar D^2 G=0\,.
\label{prop-U}
\ee
These constraints mean that $G$ is a linear superfield.
There are also the following useful relations among the superfield
strengths
\bea
D_{(\alpha} W_{\beta)}-\bar D_{(\alpha} \bar W_{\beta)}
 &=&\varepsilon^{mnp}(\gamma_p)_{\alpha\beta}{\bf F}_{mn}\,,
\label{id1}\\
D_{(\alpha} W_{\beta)}+\bar D_{(\alpha}\bar W_{\beta)}
&=&-2i\partial_{\alpha\beta} G\,.
\label{id2}
\eea

In the Abelian case the gauge connections for covariant spinor
derivatives in (\ref{covD}) can be expressed in terms of one real
gauge superfield $V$,
\be
\nabla_\alpha=e^{-2V}D_\alpha e^{2V}=D_\alpha+2D_\alpha V\,,\qquad
\bar\nabla_\alpha=\bar D_\alpha\,.
\label{40}
\ee
As a consequence of the algebra (\ref{alg1},\ref{alg2}), the
superfield strengths are given by
\be
G=\frac i2\bar D^\alpha D_\alpha V\,,\quad
W_\alpha=-\frac i4 \bar D^2D_\alpha V\,,\quad
\bar W_\alpha =-\frac i4 D^2\bar D_\alpha V\,.
\label{U-V}
\ee
Substituting (\ref{V}) into (\ref{U-V}) we find the component
structure of the superfield strengths, in particular,
\be
G=-\phi+\theta^\alpha\bar\lambda_\alpha
-\bar\theta^\alpha\lambda_\alpha
+\frac12\theta^\alpha\bar\theta^\beta f_{\alpha\beta}
-2i \theta^\alpha\bar\theta_\alpha D+\ldots,
\label{U-comp}
\ee
where $f_{\alpha\beta}=\partial_\alpha^\rho A_{\beta\rho}+\partial_\beta^\rho
A_{\alpha\rho}$ and dots stand for the terms with derivatives of
the fields.

Now we specify the constraints on the background gauge superfield
under considerations:
\begin{itemize}
\item[i)] The gauge superfield obeys the $\cN=2$ supersymmetric free Maxwell equations,
\be
D^\alpha W_\alpha=0\,,\qquad \bar D^\alpha \bar W_\alpha=0\,.
\label{on-shell}
\ee
\item[ii)] Within the derivative expansion of the effective action we
look for the leading terms without space-time derivatives of the
gauge superfields. Such a long-wave approximation
 is effectively taken into account by considering the constant
background,
\be
\partial_m G=0\,,\quad
\partial_m W_\alpha=0\,,\quad
\partial_m \bar W_\alpha=0\,.
\label{const-field}
\ee
\end{itemize}
This approximation suffices to study the
Euler-Heisenberg-type effective action which is induced by the
$\cN=2$ supersymmetric quantum matter fields.

\subsection{Superconformal invariance and the effective action}

In this subsection we analyse the general structure of the
effective action in the model (\ref{S})
employing the constraints imposed by the
gauge and superconformal invariance.
Similar analysis for the $\cN=2$, $d=4$ superconformal
models \cite{BKT} appeared very useful because it helped to
construct an off-shell extension of the terms in the gauge
superfield effective action computed in the on-shell
approximation. Here we will follow similar lines using the
realization of the superconformal group in the $\cN=2$, $d=3$
superspace developed in \cite{Park} which is a three-dimensional
extension of the general method described in \cite{bookBK}.

In general, the effective Lagrangian depends on the gauge
superfield $V$, its superfield strengths $G$, $W_\alpha$, $\bar
W_\alpha$ and their derivatives. The only gauge invariant term
with explicit dependence on the gauge superfield $V$ and which cannot
be rewritten in terms of the superfield strengths is the
Chern-Simons term \cite{ZP,NG,Ivanov92},
\be
S_{\rm CS}=\frac{k}{2\pi}\int d^3x d^4\theta\,
VG
=\frac{k}{2\pi}\int d^3x(\frac12\varepsilon^{mnp}
A_m \partial_n A_p+i\lambda^\alpha\bar\lambda_\alpha-2\phi D)\,,
\label{CS}
\ee
where $k$ is the Chern-Simons level. All other terms
in the effective Lagrangian depend only
on the superfield strengths and their derivatives.

Recall that we restricted ourself
to the long-wave approximation (\ref{const-field}) which means
that we omit all terms with space-time derivatives of superfields, but
the covariant spinor derivatives can appear in the effective
Lagrangian. In this approximation there is very limited number
of building blocks, i.e., the
superfield combinations which the effective action can depend on.
First of all, it depends on the
superfield strength $G$ as well as on $W_\alpha$ and $\bar
W_\alpha$ which involve first covariant spinor
derivatives of $G$, (\ref{W(U)}). Next, there are the
objects with two covariant spinor derivatives of $G$,
\be
N_{\alpha\beta}\equiv D_{(\alpha}W_{\beta)}\,,\qquad
\bar N_{\alpha\beta}\equiv -(N_{\alpha\beta})^*=\bar D_{(\alpha}\bar W_{\beta)}\,.
\label{N}
\ee
Note that it is sufficient to consider the objects (\ref{N}) with
symmetryzed spinor indices since
$D_{[\alpha}W_{\beta]}=\frac12\varepsilon_{\alpha\beta}D^\gamma W_\gamma
=0$ for the considered background (\ref{on-shell}).
Note also that owing to the identity (\ref{id2}), $\bar
N_{\alpha\beta}$ coincides with $N_{\alpha\beta}$ up to a sign,
\be
N_{\alpha\beta}=-\bar N_{\alpha\beta}\,,
\ee
when $\partial_m U=0$.
Finally, it is clear that any further spinor derivatives of the superfield
strengths vanish in the long-wave approximation (\ref{const-field}), e.g.,
\be
\bar D_\alpha D_\beta
W_\gamma=-2i\partial_{\alpha\beta}W_\gamma=0\,,\qquad
D^2W_\alpha=-4i\partial_{\alpha\beta}\bar W^\beta=0\,.
\ee
We conclude that the general structure of the
gauge invariant effective action is given by
\be
\Gamma_{\cN=2}=\int d^3x d^4\theta[c_0 VG+{\cal L}_{\rm eff}(G,W_\alpha,\bar W_\alpha,
N_{\alpha\beta})]\,,
\label{Geff}
\ee
where $c_0$ is an arbitrary coefficient and
${\cal L}_{\rm eff}$ is an effective Lagrangian being a real scalar
superfield. Further restrictions on the structure of the function
${\cal L}_{\rm eff}$ come from the requirement of the superconformal
invariance.

As a warming up exercise we check the superconformal invariance of
the classical action (\ref{S}). Indeed, using the explicit realization of the
superconformal group in the $\cN=2$ superspace given in the
Appendix B, we consider the superconformal transformations of the
gauge and matter superfields,
\be
\delta_{\rm sc} V = \xi V\,,\quad
\delta_{\rm sc} Q=(\sigma/2+\xi)Q\,,\quad
\delta_{\rm sc} \bar Q=(\bar\sigma/2+\xi)\bar Q\,,
\label{delta-V}
\ee
where $\xi$ is a superconformal Killing vector (\ref{xi}) and
$\sigma$, $\bar\sigma$ are (anti)chiral superfields constructed from the
parameters of the superconformal transformations,
(\ref{rho},\ref{bar-rho}). The superconformal variation of the
Lagrangian in (\ref{S}) is
\be
\delta_{\rm sc}(\bar Qe^{2V}Q)=(\rho+\xi)(\bar Qe^{2V}Q)\,,
\label{deltaL}
\ee
where $\rho=\frac12(\sigma+\bar\sigma)$ is given in (\ref{sigma}).
Equation (\ref{deltaL})
shows that $\bar Qe^{2V}Q$ is a quasi-primary scalar superfield with
conformal weight $l=+1$. Hence, according to (\ref{inv-act}), the
action (\ref{S}) is invariant under the superconformal
transformations, $\delta_{\rm sc} S_{\cN=2}=0$.

Of course, the Chern-Simons action (\ref{CS}) is superconformal
as well.
To show this, we derive the superconformal transformation of the superfield
strength $G$ with the help of (\ref{U-V},\ref{delta-V}) and (\ref{206},\ref{207}),
\be
\delta_{\rm sc} G=(\rho+\xi)G\,,
\label{deltaU}
\ee
i.e., $G$ is a quasi-primary superfield. Using (\ref{inv-act}) we
immediately find
\be
\delta_{\rm sc}S_{\rm CS}=\frac k{2\pi}\int d^3x d^4\theta
(\rho+\xi)VG=0\,.
\ee
Hence, the superconformal invariance imposes only constraints on
the function ${\cal L}_{\rm eff}$ in (\ref{Geff}).

In general, the effective Lagrangian contains the effective potential
term ${\cal F}(G)$,
\be
{\cal L}_{\rm eff}={\cal F}(G)+\tilde{\cal L}_{\rm eff}(G,W_\alpha,\bar W_\alpha,
N_{\alpha\beta})\,,
\label{F+H}
\ee
where ${\cal F}(G)$ is a holomorphic function of $G$ only while
$\tilde {\cal L}_{\rm eff}$ takes into account the superfield strength with
covariant spinor derivatives.
The superconformal invariance restricts the form of the effective potential
${\cal F}(G)$ uniquely, up to a constant. Indeed, the
general condition of superconformal invariance (\ref{inv-act1})
applied to the effective potential reads
\be
\delta_{\rm sc}{\cal F}(G)
=(\rho+\xi){\cal F}(G)
+\sigma {\cal K}(G)+\bar\sigma \bar{\cal K}(G)\,,
\label{eqL}
\ee
where the function ${\cal K}(G)$ should be linear,
\be
D^2 \bar{\cal K}(G)=\bar D^2 {\cal K}(G)=0\quad\Rightarrow\quad
{\cal K}(G)=\alpha+\beta G\,,\quad
\bar{\cal K}(G)=\bar\alpha+\bar\beta G\,,
\ee
with $\alpha$ and $\beta$ being some (complex) constants. Up to the terms
vanishing under integral over full $\cN=2$ superspace, the general
solution of (\ref{eqL}) is given by
\footnote{To be more precise, we have to use
a dimensionless combination $G/\mu$ under the logarithm in (\ref{GlnG})
where $\mu$ is some scale. However, this
parameter is spurious and drops out completely in the component
field formulation. Therefore we omit this parameter everywhere for brevity.}
\be
{\cal F}(G)=c_1 G\ln G\,,
\label{GlnG}
\ee
where $c_1$ is some constant. This effective potential is
responsible for a superconformal generalization of the Maxwell
term in its component decomposition,
\be
\int d^3x d^4\theta\,G\ln G=\frac18\int d^3x\frac1\phi
F^{mn}F_{mn}+\ldots,
\label{F2}
\ee
where dots stand for other component terms.
Note that the Lagrangian (\ref{GlnG}) being considered in the $\cN=1$, $d=4$
superspace is responsible for the classical action of the improved tensor
 multiplet model \cite{bookBK}.

It is much more difficult to make general analysis of the admissible
form of the function $\tilde{\cal L}_{\rm eff}$ in (\ref{F+H}) subject
to the superconformal invariance of the corresponding action. The problem is that the superfields
$W_\alpha$, $\bar W_\alpha$ and $N_{\alpha\beta}$ are not quasi-primary,
e.g.,
\be
\label{dW}
\delta_{\rm sc} W_\alpha=(\frac12\rho+\sigma+\xi)W_\alpha
+\omega_{\alpha\beta}W^\beta+(\bar
D_\alpha\rho)G\,,
\ee
where $\omega_{\alpha\beta}=\bar D_{(\alpha}\bar\xi_{\beta)}
=-D_{(\alpha}\xi_{\beta)}$ are the parameters of `local' Lorentz
transformations.
Equation (\ref{dW}) shows that $W_\alpha$ transforms
inhomogeneously because of the last term in (\ref{dW}).
This is a new
feature of three-dimensional supergauge models as compared
to the $\cN=1$, $d=4$ ones in which the superfield
strengths are chiral quasi-primary, \cite{bookBK,Osborn,Park2}.
Therefore the superfields $W_\alpha$ and $\bar W_\alpha$ are
rather inconvenient for constructing superconformal actions
and we are forced to introduce the following quasi-primary
superfields
\footnote{In principle, the coefficients in (\ref{Psin}) are
arbitrary, but we fix them in such a way that the effective action
(\ref{t3}) computed in the next section gets simple form.}
\be
\Psi=\frac iG\bar D^\alpha D_\alpha \ln G\,,\qquad
\Omega^2=\frac18(\frac 1G\bar D^\alpha D_\alpha)^2\ln G\,.
\label{Psin}
\ee
Indeed, using (\ref{deltaU}) and the relations
(\ref{206},\ref{207}) one can readily
check that both these superfields are quasi-primary with zeroth scaling dimension,
\be
\delta_{\rm sc} \Psi=\xi \Psi\,,\qquad
\delta_{\rm sc} \Omega^2=\xi\Omega^2\,.
\label{delta-psi}
\ee
This allows us to construct a superconformal action with these
superfields,
\be
S_1=\int d^3x d^4\theta\, G\,{\cal U}(\Psi,\Omega^2)\,,\qquad
\delta_{\rm sc}S_1=0\,,
\label{S1_}
\ee
where ${\cal U}(\Psi,\Omega^2)$ is an arbitrary function.

Neither the gauge invariance nor the superconformal symmetry
impose any restrictions on possible form of the function
${\cal U}(\Psi,\Omega^2)$ in (\ref{S1_}). However, for the
background gauge superfield under considerations
(\ref{on-shell},\ref{const-field}) the form of this functions can
be further reduced. Indeed, for such a background
there are the following equivalent representations for $\Psi$ and
$\Omega^2$,
\bea
\Psi&=&-i\frac{W^\alpha\bar W_\alpha}{G^3}
\label{Psi}\,,\\
\Omega^2&=&\frac18\frac{N^\alpha_\beta N_\alpha^\beta}{G^4}
+\frac34\frac{ N^{\alpha\beta}W_\alpha\bar W_\beta}{G^5}
+\frac34\frac{W^2\bar W^2}{G^6}\,.
\label{Omega2}
\eea
Owing to the odd statistics of superfield strengths $W_\alpha$ and $\bar
W_\alpha$, the power expansion of ${\cal U}(\Psi,\Omega^2)$ over
$\Psi$ terminates at the second order,
\be
{\cal U}(\Psi,\Omega^2)={\cal U}_0(\Omega^2)
+\Psi{\cal U}_1(\Omega^2)+\Psi^2{\cal U}_2(\Omega^2)\,.
\label{cU}
\ee
Under the integral over $\cN=2$ superspace the first two terms in
the rhs of (\ref{cU}) can be brought to the form of the last term,
\be
\int d^3x d^4\theta\,G[{\cal U}_0(\Omega^2)
+\Psi{\cal U}_1(\Omega^2)]
=\int d^3x d^4\theta\,G\Psi^2\tilde{\cal U}_2(\Omega^2)\,,
\label{cU1}
\ee
where $\tilde {\cal U}_2$ is some function. Indeed, to check
(\ref{cU1}) one has to take the covariant spinor derivatives from
$N_{\alpha\beta}=D_\alpha W_\beta$ in (\ref{Omega2}) and integrate them by parts.
Note that these derivatives hit only the superfield $G$ but not
$N_{\alpha\beta}$ because of the restrictions
(\ref{const-field}). Doing so, one can accumulate the
factor $W^2\bar W^2$ in the nominator resulting in $\Psi^2$
according to (\ref{Psi}) while the remaining factors can be
represented by some function $\tilde {\cal U}_2(\Omega^2)$.

These considerations show that in the long-wave approximation the
superconformal action (\ref{S1_}) simplifies
\be
S_1=\int d^3x d^4\theta\, G\Psi^2{\cal H}(\Omega^2)\,,
\label{S2__}
\ee
such that it is described by a single function ${\cal H}(\Omega^2)$ of one
real variable. There are no any more constraints on the
form of this function. This function will be computed explicitly
in the next subsection, but in general, it is represented by a power
series ${\cal H}(\Omega^2) =\sum_{n=0}^\infty a_n
\Omega^{2n}$ with some coefficients $a_n$. The action
(\ref{S2__}) contains the following terms in its component
decomposition
\be
S_1=\sum_{n=0}^\infty a_n
\int d^3xd^4\theta\frac{W^2\bar W^2}{2G^5}
\left(
\frac18\frac{N^\alpha_\beta N_\alpha^\beta}{G^4}
\right)^n
=\sum_{n=0}^\infty \left(-\frac18\right)^{n+1} a_n
\int d^3x\frac{F^{4+2n}}{\phi^{5+4n}}+\ldots\,.
\label{F2n}
\ee

Summing up all together, we conclude that the general form of the
superconformal effective action in the long-wave approximation is given by
\be
\Gamma_{\cN=2}=\Gamma_{\rm CS}+\Gamma_{\rm Maxweel}+\Gamma_{\rm
higher}\,,
\label{Gamma-full}
\ee
where
\bea
\Gamma_{\rm CS}&=&c_0\int d^3x d^4\theta\, VG\,,
\label{GCS}\\
\Gamma_{\rm Maxwell}&=&c_1\int d^3x d^4\theta\, G\ln G\,,
\label{GM}\\
\Gamma_{\rm higher}&=&\int d^3x d^4\theta\, G\Psi^2 {\cal
H}(\Omega^2)\,.
\label{GH}
\eea
In components, this action contain the Chern-Simons term
(\ref{CS}), the Maxwell $F^2$ term (\ref{F2}) and all higher order
terms $F^{2n}$ with $n\geq 2$ which are written down in
(\ref{F2n}). The undefined coefficients $c_0$, $c_1$ and the
arbitrary function ${\cal H}$ will be found in the next section by
explicit quantum computations.

In conclusion of this subsection we comment on the uniqueness of
the form of the superconformal action (\ref{Gamma-full}). The
Chern-Simons and the Maxwell terms in this action are fixed by
the gauge and superconformal invariance uniquely, up to the
coefficients, but the form of the last term $\Gamma_{\rm
higher}$ is not unique. Indeed, we used the ansatz (\ref{S1_}) which
involves only two quasi-primary superfields (\ref{Psin}), but the
other ans\"atze are also possible. In particular, there is a
consequence of descendant quasi-primary superfields for
(\ref{Psin}) given by
\be
\Psi_n=(\frac iG\bar D^\alpha D_\alpha )^n\ln G\,,\quad
\delta \Psi_n=\xi\Psi_n\,,\quad
n=3,4,5,\ldots,
\label{Psi-high}
\ee
which can also be used for constructing superconformal invariants
in the $\cN=2$ superspace,
\be
\tilde S=\int d^3x d^4\theta\, G\, {\cal A}(
\Psi,\Omega^2,\Psi_3,\Psi_4,\ldots,\Psi_n,\ldots)\,,
\qquad
\delta\tilde S=0\,,
\label{S2_}
\ee
where ${\cal A}$ is some function. However, these
superfields (\ref{Psi-high}) are necessary only for describing the
terms involving space-time derivatives of the gauge superfields or
which are proportional to the free Maxwell equations since in the
long-wave approximation (\ref{const-field}) the action (\ref{S2_})
can be brought to the form (\ref{S2__}) by similar manipulations as
in (\ref{cU1}).

\setcounter{equation}{0}
\section{Perturbative computations of $\cN=2$ and $\cN=4$
effective actions}
\subsection{Low-energy effective action for $\cN=2$ gauge superfield}
The four-dimensional $\cN=1$ and $\cN=2$ supersymmetric Euler-Heisenberg
effective actions were studied in \cite{d4eff,BKT} at one loop. The
two-loop refining of these results was given in \cite{Kuz03} and
\cite{Kuz07} owing to the powerful covariant perturbation theory
in superspace elaborated in \cite{Kuz03-cov}. Here we apply some
of the methods developed in \cite{d4eff,Kuz03,Kuz07} for studying the
structure of the effective action in three-dimensional model of
chiral superfield interacting with the background gauge
superfield.

The effective action for the model (\ref{S}) can be divided into parity odd and
parity even parts,
\be
\Gamma_{\cN=2}=\Gamma_{\rm odd}+\Gamma_{\rm even}\,.
\label{GammaN2}
\ee
As soon as the classical action (\ref{S}) is parity even, the
appearance of the odd part in the effective action can be only due
to the parity anomaly which is studied in details in \cite{Redlich,Semenoff}
and reviewed in \cite{Dunne}. The anomaly appears owing to the
regularization of infrared divergent momentum integrals and
yields the term proportional to the Chern-Simons action
$
\Gamma_{\rm odd}\propto S_{\rm CS}$.
We will compute $\Gamma_{\rm odd}$ in the end of this subsection while now
we concentrate on $\Gamma_{\rm even}$.

For $Q$ and $\bar Q$ it is convenient to introduce the covariantly
(anti)chiral superfields \cite{GGRS},
\be
{\cal Q}=Q\,,\qquad
\bar{\cal Q}=\bar Q e^{2V}\,,
\ee
which are annihilated by the gauge covariant derivatives (\ref{40}),
\be
\bar\nabla_\alpha {\cal Q}=0\,,\qquad
\nabla_\alpha\bar{\cal Q}=0\,.
\ee
In terms of these superfields the action (\ref{S}) is simply
\be
S_{\cN=2}=-\int d^3xd^4\theta\,\bar{\cal Q}{\cal Q}\,.
\ee
The matrix of second variational derivatives of this action
is given by
\be
H=
\left(
\begin{array}{cc}
\frac{\delta^2S}{\delta{\cal Q}(z)\delta{\cal Q}(z')}&
\frac{\delta^2S}{\delta{\cal Q}(z)\delta\bar{\cal Q}(z')}\\
\frac{\delta^2S}{\delta\bar{\cal Q}(z)\delta{\cal Q}(z')} &
\frac{\delta^2S}{\delta\bar{\cal Q}(z)\delta\bar{\cal Q}(z')}
\end{array}
\right)=
\left(
\begin{array}{cc}
0 & \frac14\bar \nabla^2 \delta_-(z,z') \\
\frac14 \nabla^2 \delta_+(z,z') &0
\end{array}
\right),
\label{H}
\ee
where $\delta_+$ and $\delta_-$ are covariantly (anti)chiral
delta-functions,
\be
\delta_{+}(z,z')=-\frac14\bar \nabla^2\delta^7(z-z')\,,
\qquad
\delta_{-}(z,z')=-\frac14\nabla^2\delta^7(z-z')\,.
\ee
Here $\delta^7(z-z')$ is the delta-function in full $\cN=2$
superspace.

The matrix (\ref{H}) leads to the following one-loop effective action,
\be
\Gamma_{\rm even}=\frac i2\Tr\ln H=\frac i4 \Tr\ln H^2
=\frac i4\Tr\ln\left(
\begin{array}{cc}
\frac1{16} \bar \nabla^2
 \nabla^2 \delta_+(z,z')&0\\
0&\frac1{16}  \nabla^2 \bar \nabla^2 \delta_-(z,z')
\end{array}
\right).
\label{Gamma1}
\ee
Introducing the covariant (anti)chiral
d'Alembertians
\be
\square_+=\frac1{16}\bar\nabla^2\nabla^2\,,\qquad
\square_-=\frac1{16}\nabla^2\bar\nabla^2\,,
\ee
the effective action (\ref{Gamma1}) can be rewritten as
\be
\Gamma_{\rm even}=\frac i4\Tr_+\ln\square_++\frac i4\Tr_-\ln\square_-\,,
\label{gamma1}
\ee
where $\Tr_+$ and $\Tr_-$ denote the functional traces of the
corresponding operators in the chiral and antichiral superspaces,
respectively. The operators $\square_+$ and $\square_-$ acting on the covariantly
(anti)chiral superfields have the following representations
\bea
\square_+&=&\nabla^m\nabla_m+G^2+\frac i2(D^\alpha W_\alpha)
+iW^\alpha\nabla_\alpha\,,
\label{box+}\\
\square_-&=&\nabla^m\nabla_m+G^2-\frac i2(\bar D^\alpha \bar W_\alpha)
-i\bar W^\alpha\bar\nabla_\alpha\,.
\label{box-}
\eea
The terms $D^\alpha W_\alpha$ and $\bar D^\alpha\bar
W_\alpha$ in (\ref{box+}) and (\ref{box-}) can be omitted as soon
as we consider the special background, (\ref{on-shell}).
Then the operators (\ref{box+}) and (\ref{box-}) obey the following
important properties
\be
\nabla^2\square_+=\square_-\nabla^2\,,\qquad
\bar\nabla^2\square_-=\square_+\bar\nabla^2\,.
\ee
There are covariantly (anti)chiral Greens functions $G_+$ and
$G_-$ for these operators,
\be
\square_+ G_+(z,z')=-\delta_+(z,z')\,,\qquad
\square_- G_-(z,z')=-\delta_-(z,z')\,.
\label{chiralG}
\ee
These Greens functions can be represented by their heat kernels,
\be
G_\pm(z,z')=i\int_0^\infty ds\, K_\pm(z,z'|s)e^{-\epsilon s}\,,
\qquad
\epsilon\to+0\,.
\ee
Further we will omit the factor $e^{-\epsilon s}$ in the integrals
over the proper time $s$ for brevity assuming the limit
$\epsilon\to+0$ after calculating the integrals.
In terms of the chiral heat kernel the effective action (\ref{gamma1})
reads
\be
\Gamma_{\rm even}=-\frac i4\int_0^\infty\frac{ds}{s}\Tr_+K_+(s)+c.c\,.
\label{Gam}
\ee
As a result, the problem of computing the effective action
is reduced to finding the coincidence limit of the chiral heat kernel,
\be
\Tr_+ K_+(s)=\int d^3x d^2\theta\,K_+(z,z|s)\,.
\ee

Let us introduce the operator
\be
\square_{\rm v}=\nabla^m\nabla_m+G^2+iW^\alpha \nabla_\alpha-i\bar W^\alpha
\bar \nabla_\alpha\,,
\ee
with the Greens function $G_{\rm v}$ and associated heat kernel $K_{\rm v}$,
\be
\square_{\rm v} G_{\rm v}(z,z')=-\delta^7(z,z')\,,\qquad
G_{\rm v}(z,z')=i\int_0^\infty ds\, K_{\rm v}(z,z'|s)\,.
\label{Gv}
\ee
For the special background under considerations,
 $D^\alpha W_\alpha=\bar D^\alpha\bar W_\alpha=0$,
this operator has the following important properties
\be
\nabla^2\square_+=\nabla^2\square_{\rm v}=\square_{\rm v}\nabla^2\,,\qquad
\bar\nabla^2\square_-=\bar\nabla^2\square_{\rm v}=\square_{\rm v}\bar\nabla^2\,,
\ee
which are used to relate the (anti)chiral Greens functions (\ref{chiralG}) with $G_{\rm v}$,
\be
G_+(z,z')=-\frac14\bar \nabla^2 G_{\rm v}(z,z')\,,\qquad
G_-(z,z')=-\frac14\nabla^2 G_{\rm v}(z,z')\,,
\label{relG}
\ee
as well as the corresponding heat kernels,
\be
K_+(z,z'|s)=-\frac14 \bar \nabla^2 K_{\rm v}(z,z'|s)\,,\qquad
K_-(z,z'|s)=-\frac14 \nabla^2 K_{\rm v}(z,z'|s)\,.
\label{K+Kv}
\ee
Therefore it is sufficient to study the heat kernel $K_{\rm v}$ while
the (anti)chiral ones are deduced from $K_{\rm v}$ by (\ref{K+Kv}).

The heat kernel $K_{\rm v}$ can be represented as
\be
K_{\rm v}(z,z'|s)=e^{is(\nabla^m\nabla_m+G^2+iW^\alpha\nabla_\alpha
-i\bar W^\alpha\bar\nabla_\alpha)}\delta^7(z-z')\,.
\label{Kv}
\ee
For the constant field background
(\ref{on-shell},\ref{const-field})
there are the following identities
\be
[\nabla_m,W^\alpha\nabla_\alpha
-\bar W^\alpha\bar\nabla_\alpha]=0\,,\qquad
[W^\alpha\nabla_\alpha
-\bar W^\alpha\bar\nabla_\alpha,G]=0\,,
\ee
which allow us to factorize the exponent in (\ref{Kv}),
\be
K_{\rm v}(z,z'|s)=e^{isG^2}e^{is(iW^\alpha\nabla_\alpha
-i\bar W^\alpha\bar\nabla_\alpha)}e^{is(\nabla^m\nabla_m)}\delta^7(z-z')
\equiv e^{isG^2}{\cal O}(s)\tilde K(z,z'|s)\,,
\label{Kv1}
\ee
where
\be
{\cal O}(s)=e^{s(\bar W^\alpha\bar\nabla_\alpha-W^\alpha\nabla_\alpha)}
\label{O}
\ee
and the reduced kernel $\tilde K(z,z'|s)$ solves the equation
\be
(i\frac d{ds}+\nabla^m\nabla_m)\tilde K(z,z'|s)=0\,,\qquad
\lim_{s\to0}\tilde K(z,z'|s)=\delta^7(z-z')\,.
\label{tildeK}
\ee

Let us consider the following representation
for the delta-function in full superspace
\footnote{More generally, one has to insert a parallel displacement
operator on the right in (\ref{covdelta}) as well as to further expressions for heat kernels
to provide their gauge covariance \cite{Kuz03-cov}. However, this is not necessary
for the one-loop computations since in the limit of coincident
points the covariant derivatives of the parallel displacement
operator vanish, \cite{Kuz03,Kuz07,Kuz03-cov}.}
\be
\delta^7(z-z')=\int
\frac{d^3k}{(2\pi)^3}e^{ik^m\zeta_m}\zeta^2\bar\zeta^2 \,,
\label{covdelta}
\ee
where $\zeta^A$ is $\cN=2$ supersymmetric interval,
\be
\zeta^A=\left\{
\begin{array}{rcl}
\zeta^{\alpha\beta}&=&(x-x')^{\alpha\beta}
-2i(\theta-\theta')^{(\alpha}\bar\theta'^{\beta)}
+2i\theta'^{(\alpha}(\bar\theta-\bar\theta')^{\beta)}\,,\\
\zeta^\alpha&=&(\theta-\theta')^\alpha\,,\\
\bar\zeta^\alpha&=&(\bar\theta-\bar\theta')^\alpha\,.
\end{array}
\right.
\ee
Using (\ref{tildeK}) and (\ref{covdelta}) we arrive at the
following representation for the heat kernel $\tilde K$,
\be
\tilde K(z,z'|s)=\int \frac{d^3k}{(2\pi)^3}
e^{ik^n\zeta_n}e^{is(\nabla^m+ik^m)(\nabla_m+ik_m)}\zeta^2\bar\zeta^2
\,.
\label{tildeK1}
\ee
The integration over $d^3k$ in (\ref{tildeK1}) can be explicitly
done, see \cite{Kuz03-cov} for the details of similar
computations in four-dimensional case,
\be
\tilde K(z,z'|s)=\frac1{8(i\pi s)^{3/2}} \sqrt{\det
\left(\frac{2s{\bf F}}{1-e^{-2s{\bf F}}}\right)}
e^{\frac i4({\bf F}\coth s{\bf F})_{mn}\zeta^m\zeta^n}
\zeta^2\bar\zeta^2 \,.
\label{tildeK2}
\ee
The determinant in (\ref{tildeK2}) is over the Lorentz indices of
the matrix ${\bf F}_m{}^n$ introduced in (\ref{algebra}). This
determinant can be explicitly evaluated (see \cite{Redlich} for
analogous computations in the non-supersymmetric
three-dimensional electrodynamics)
\be
\sqrt{\det\left(\frac{2s{\bf F}}{1-e^{-2s\bf F}}\right)}
=\frac{sB}{\sinh (sB)}\,,
\ee
where
\be
B^2=\frac12N_\alpha^\beta N_\beta^\alpha\,,
\quad
N_{\alpha\beta}=D_{(\alpha}W_{\beta)}\,,\quad
\bar N_{\alpha\beta}=\bar D_{(\alpha}\bar W_{\beta)}\,.
\label{B2}
\ee
Here the identity (\ref{id1}) has been used.

Now we return to the computation of the heat kernel $K_{\rm v}$ which
is expressed in terms of $\tilde K$ as in (\ref{Kv1}). For this purpose
we need to push the operator ${\cal O}(s)$ through the components
of the superinterval $\zeta^A$ in (\ref{tildeK2}). Using the
identities
\bea
W^\alpha(s)&\equiv&{\cal O}(s)W^\alpha {\cal O}(-s)
=W^\beta (e^{-sN})_\beta{}^\alpha\,,\\
\zeta^\alpha(s)&\equiv&{\cal O}(s)\zeta^\alpha {\cal O}(-s)
= \zeta^\alpha+W^\beta
((e^{-sN}-1)N^{-1})_\beta{}^\alpha\,,\\
\zeta_{\alpha\beta}(s)&\equiv&{\cal O}(s)\zeta_{\alpha\beta}{\cal O}(-s)
=\zeta_{\alpha\beta}-2i\int_0^sdt(W_{(\alpha}(t)\bar\zeta_{\beta)}(t)
+\bar W_{(\alpha}(t)\zeta_{\beta)}(t))\,,
\eea
we arrive at the following final expression for $K_{\rm v}$
\be
K_{\rm v}(z,z'|s)=\frac1{8(i\pi s)^{3/2}}
\frac{sB}{
\sinh(sB)}
e^{isG^2}
e^{\frac i4({\bf F}\coth s{\bf F})_{mn}\zeta^m(s)\zeta^n(s)}
\zeta^2(s)\bar\zeta^2(s) \,.
\label{Kv-final}
\ee

Recall that we need the chiral heat kernel $K_+$ for the effective
action (\ref{Gam}), which is related to $K_{\rm v}$ by (\ref{K+Kv}).
At coincident points, $z=z'$, it is easy to argue that the
operator $\bar\nabla^2$ in (\ref{K+Kv}) hits only $\bar\zeta^2(s)$,
\be
-\frac14\bar\nabla^2\bar \zeta^2(s)=1\,.
\ee
Finally, using the identity
\be
\zeta^2(s)|_{z=z'}=s^2 W^2\frac{\sinh^2\frac{sB}{2}}{(sB/2)^2}\,,
\ee
we get
\be
K_+(s)= K_+(z,z|s)
=\frac{1}{8(i\pi s)^{3/2}}s^2 W^2 e^{isG^2}
\frac{\tanh (sB/2)}{sB/2}\,.
\ee
The corresponding one-loop effective action (\ref{Gam}) reads
\be
\Gamma_{\rm even}=-\frac1{32\pi}\int d^3x d^2\theta
\int_0^\infty \frac{ds}{\sqrt{i\pi
s}}W^2e^{isG^2}\frac{\tanh(sB/2)}{sB/2}+c.c.\,,
\label{Gamma-fin}
\ee
where $B$ is given by (\ref{B2}),
$B^2=\frac12 D_\alpha W^\beta D_\beta W^\alpha$.
Finally, we rewrite (\ref{Gamma-fin}) in the full $\cN=2$
superspace,
\be
\Gamma_{\rm even}=\frac1{4\pi}\int d^3xd^4\theta\left[G
\ln G
+\frac14\int_0^\infty\frac{ds}{\sqrt{i\pi s}}e^{isG^2}
\frac{W^2\bar W^2}{B^2}\left(
\frac{\tanh(sB/2)}{sB/2}-1
\right)
\right].
\label{Gamma-fin1}
\ee

The superfield strength $G$ in (\ref{Gamma-fin}) serves as an
effective massive regularizator for infrared divergencies. In
fact, giving a non-zero vev $\langle G\rangle\ne0$ generates a
mass for the matter superfield which is equal to the central
charge of the $\cN=2$ superalgebra. In other words, we derived the
effective action (\ref{Gamma-fin}) in the Coulomb branch of the
$\cN=2$ supergauge theory. Alternatively, one can
consider standard mass term $m\int d^3xd^2\theta Q^2$, but it violates
the parity and requires more accurate considerations.
These issues were studied in details in \cite{AHKSS}.
We will consider the hypermultiplet model with the complex mass
in the next section.

Now we come back to the derivation of the parity odd part of the
effective action (\ref{GammaN2}). The reason why both $\Gamma_{\rm
odd}$ and $\Gamma_{\rm even}$ cannot be derived in the unified
procedure given above is quite similar to the non-supersymmetric
case considered in \cite{Redlich}: The Chern-Simons term formally
vanishes in the approximation of the constant fields
(\ref{const-field}), but the variation of the Chern-Simons term
with respect to the gauge superfield produces a non-vanishing
current. Therefore the Chern-Simons term in the effective action
can be obtained by integrating the variation
\be
\delta \Gamma_{\cN=2}=\int d^3x d^4\theta\, \delta V \langle J\rangle\,,
\label{gamma-var}
\ee
where $\langle J\rangle$ is the effective current,
\be
\langle J\rangle =\langle \frac{\delta S}{\delta V}\rangle
=-2\langle \bar {\cal Q}{\cal Q} \rangle\,.
\ee
The propagator $\langle \bar {\cal Q}{\cal Q} \rangle$ is
expressed in terms of the Greens function (\ref{Gv}),
\be
i\langle \bar {\cal Q}(z){\cal Q}(z') \rangle
=\frac1{16}\bar\nabla^2\nabla'{}^2 G_{\rm v}(z,z')\,.
\ee
Using the explicit form (\ref{Kv-final}) for $G_{\rm v}$, we find
\be
\langle J\rangle=
\frac i8\bar\nabla^2\nabla'{}^2 G_{\rm v}(z,z')|_{z=z'}
=-\frac14\int_0^\infty
\frac{ds}{(i\pi s)^{3/2}}\frac{sB}{\sinh(sB)}e^{isG^2}\,.
\label{J1}
\ee
The effective current (\ref{J1}) contains both finite and infrared divergent
parts. All finite contributions to the effective action are parity
even and are already taken into account in (\ref{Gamma-fin}). The parity
odd contributions arise from the divergent part which reads
\be
\langle J\rangle_{\rm div}=-\frac14\int_0^\infty
\frac{ds}{(i\pi s)^{3/2}}e^{isG^2}\,.
\label{J2}
\ee
Regularizing this integral appropriately we find
\be
\langle J\rangle_{\rm reg}=\frac{G}{2\pi}\,.
\label{J3}
\ee
Substituting this current into (\ref{gamma-var}) we obtain the odd
part of the effective action,
\be
\Gamma_{\rm odd}=\frac1{4\pi}\int d^3xd^4\theta\, VG\,.
\label{Gamma-CS}
\ee

Summing up (\ref{Gamma-fin}) with (\ref{Gamma-CS}) we get the
resulting effective action in the form
(\ref{Gamma-full}) with
\bea
\Gamma_{\rm CS}&=&\frac1{4\pi}\int d^3x d^4\theta\,VG\,,
\label{t1}\\
\Gamma_{\rm Maxwell}&=&\frac1{4\pi}\int d^3x d^4\theta
\,G\ln G\,,
\label{t2}\\
\Gamma_{\rm higher}&=&
\frac1{32\pi}\int d^3x d^4\theta\, G
\frac{\Psi^2}{\Omega^2}\int_0^\infty
\frac{dt\, e^{it}}{\sqrt{i\pi t}}
\left(\frac{\tanh(t\Omega)}{t\Omega}-1
\right).
\label{t3}
\eea
The action (\ref{t3}) is obtained from (\ref{Gamma-fin1}) by
changing to the dimensionless integration variable $t=sG^2$ and
then by rewriting it in terms of the quasi-primary superconformal
superfields (\ref{Psin}) with the help of
(\ref{Psi},\ref{Omega2}). As is demonstrated in the previous
section each of the actions (\ref{t1}), (\ref{t2}) and
(\ref{t3}) is explicitly superconformal.

Representing the effective action (\ref{Gamma-fin1}) in the
superconformal form (\ref{t2},\ref{t3}) allows us to relax the on-shell constraint
(\ref{on-shell}). Indeed, there are infinitely many ways of
complementing the effective action (\ref{Gamma-fin1}) by the terms vanishing on
the classical equations of motion, but the superconformal
invariance fixes this freedom and gives the unique answer
(\ref{t2},\ref{t3}) for such an action. Therefore we conclude that
(\ref{t1},\ref{t2},\ref{t3}) are correct off-shell contributions
to the low-energy effective action of the chiral superfield
interacting with the background gauge superfield. These
conclusions are completely analogous to the ones in \cite{BKT}
for the four-dimensional $\cN=2$ superconformal theories.

In principle, one can think that the superconformal invariance
allows one to go beyond the long-wave approximation
(\ref{const-field}), but it is not completely true because when
the space-time derivatives are taken into account the terms (\ref{t2}) and
(\ref{t3}) in the effective action may be corrected by some
contributions involving the higher-order superconformal invariants
(\ref{S2_}). The analysis of contributions to the effective action
with space-time derivatives is a hard task and therefore we
restrict ourself to the long-wave approximation
(\ref{const-field}).

An interesting feature of the three-dimensional theory is that the
proper time integral in (\ref{Gamma-fin})
can be expressed in terms of special functions. In particular,
for real $B$
(constant electric field) this integral is represented by
the following combination of generalized Riemann zeta functions,%
\footnote{There is a definition of the generalized Riemann zeta
function,
$\zeta(s,q)=\sum_{n=0}^\infty (q+n)^{-s}$, valid for $\mbox{Re}(s)>1$,
$\mbox{Re}(q)>0$, but it can be analytically continued for other
values of arguments. This function is also referred to
as Hurwitz zeta function (see, e.g., \cite{Math}).}
\be
\Gamma_{\rm even}=\frac{1-i}{4\pi}\int d^3x d^4\theta\frac{ W^2\bar W^2}{ B^{5/2}}
\left[
\zeta(-\frac12,-\frac{iG^2}{2B})
-2\zeta(-\frac12,\frac12-\frac{iG^2}{2B})
+\zeta(-\frac12,1-\frac{iG^2}{2B})
\right].
\label{G-zeta}
\ee
This representation allows us to consider strong electric field
background, $B\gg1$,
\be
\Gamma_{\rm even}=\frac{1-i}{4\pi}\int d^3x d^4\theta\frac{ W^2\bar W^2}{ B^{5/2}}
\left[(\sqrt2-4)\zeta(-1/2)+O(B^{-1})
\right].
\ee
The non-vanishing imaginary part of the effective action shows
the vacuum instability for strong electric field.
For imaginary $B$ (constant magnetic field) one can replace
$B\to -iB$ in (\ref{G-zeta}) to see that the effective action is
real for any value of the field.

\subsection{Low-energy effective action for $\cN=4$ gauge multiplet}
The classical action for $\cN=4$, $d=3$ hypermultiplet appears by
dimensional reduction from $\cN=2$, $d=4$ hypermultiplet
which is described in $\cN=1$, $d=4$ superspace in \cite{bookBK,GGRS}.
In our case it is given by a pair of chiral superfields
$(Q_+,Q_-)$ in the $\cN=2$, $d=3$ superspace where the subscripts `$+$' and
`$-$' stress that these superfields have corresponding charges
with respect to the gauge superfield. We consider the
minimal gauge interaction of the hypermultiplet with the
$\cN=4$ vector multiplet described by the pair
$(V,\Phi)$, where $V$ is a real gauge $\cN=2$ superfield and $\Phi$
is a chiral $\cN=2$ superfield. The corresponding massless action reads
\be
S_{\cN=4}=-\int d^3xd^4\theta\left(
\bar Q_+e^{2V}Q_+ + \bar Q_- e^{-2V} Q_-  \right)
-\left( \int d^3x d^2\theta\, Q_+\Phi Q_- +c.c.\right).
\label{SN4}
\ee
The massive case can be obtained from (\ref{SN4}) by the shift
$\Phi\to\Phi+m$ with $m$ being a complex mass parameter.

It is convenient to unify the chiral superfields $Q_+$ and $Q_-$
with opposite charges to a chiral doublet $\bf Q$ \cite{Kuz03},
\be
{\bf Q}=\exp\left(i\frac \pi4 \sigma_1\right)
\left(
\begin{array}{c}
Q_+\\ Q_-
\end{array}
\right),
\ee
while for the gauge superfield $V$ and its superfield strengths we introduce
\be
{\bf V}=\sigma_2 V\,,\quad
{\bf G}=\sigma_2 G\,,\quad
{\bf W}_\alpha=\sigma_2 W_\alpha\,,\quad
\bar{\bf W}_\alpha=\sigma_2 \bar W_\alpha\,,
\label{VV}
\ee
where $\sigma_1$, $\sigma_2$, $\sigma_3$ are the Pauli matrices.
Then the action (\ref{SN4}) reads
\be
S_{\cN=4}=-\int d^3x d^4\theta\,
\bar {\bf Q}^{\rm T} e^{2\bf V}{\bf Q}
+\left(\frac i2 \int d^3x d^2\theta\, \Phi{\bf Q}^{\rm T} {\bf Q}
+c.c.\right).
\label{SN4-1}
\ee
We will also use the covariant spinor derivatives covariantized by
the matrix gauge superfield (\ref{VV}),
\be
\nabla_\alpha=D_\alpha+2D_\alpha {\bf V}\,,\qquad
\bar\nabla_\alpha=\bar D_\alpha\,,
\label{bf-nabla}
\ee
as well as the covariantly chiral superfields,
\be
\bar{\cal Q}=e^{-2\bf V}\bar{\bf Q}\,,\qquad
{\cal Q}={\bf Q}\,.
\ee
With these notations the action (\ref{SN4-1}) takes the form
\be
S_{\cN=4}=-\int d^3x d^4\theta\,\bar {\cal Q}^{\rm T} {\cal Q}
+\left( \frac i2\int d^3x d^2\theta\, \Phi{\cal Q}^{\rm T} {\cal Q}
+c.c.\right).
\label{SN4-2}
\ee

We are interested in the one-loop effective action $\Gamma_{\cN=4}[V,\Phi]$
 in the model (\ref{SN4}) which is obtained by integrating out
the charged hypermultiplet with $V$ and $\Phi$
being the background superfields. The constraints on the considered vector background
(\ref{on-shell},\ref{const-field}) should be extended by the
following constraint on $\Phi$,
\be
D_\alpha \Phi=0\,.
\label{backPhi}
\ee

Similarly as in the previous subsection, we compute the matrix of
second variational derivatives,
\be
H=
\left(
\begin{array}{cc}
\frac{\delta^2S_{\cN=4}}{\delta{\cal Q}(z)\delta{\cal Q}(z')}&
\frac{\delta^2S_{\cN=4}}{\delta{\cal Q}(z)\delta\bar{\cal Q}(z')}\\
\frac{\delta^2S_{\cN=4}}{\delta\bar{\cal Q}(z)\delta{\cal Q}(z')} &
\frac{\delta^2S_{\cN=4}}{\delta\bar{\cal Q}(z)\delta\bar{\cal Q}(z')}
\end{array}
\right)=
\left(
\begin{array}{cc}
i\Phi \delta_+(z,z') & \frac14\bar \nabla^2 \delta_-(z,z') \\
\frac14 \nabla^2 \delta_+(z,z') &i\bar\Phi\delta_-(z,z')
\end{array}
\right),
\label{HH}
\ee
where $\delta_+$ and $\delta_-$ are gauge covariant (anti)chiral
delta-functions defined with respect to the gauge covariant
derivatives (\ref{bf-nabla}). Then the one-loop effective action
reads \cite{Kuz03}
\be
\Gamma_{\cN=4}=\frac i2{\bf Tr}\ln H=
\frac i2\Tr_+\ln(\square_++\frac1{16}\bar\nabla^2\bar\Phi
\frac1{\square_-}\nabla^2\Phi)+c.c.
\label{357}
\ee
Here $\bf Tr$ takes into account not only the functional trace of
the corresponding operators, but also the matrix trace since
we deal with the matrix gauge superfield (\ref{VV}). One can
easily see that the matrix trace gives extra coefficient 2 in
(\ref{357}) as compared to (\ref{gamma1}).

For the considered background (\ref{backPhi}) the expression
(\ref{357}) simplifies,
\be
\Gamma_{\cN=4}=
\frac i2\Tr_+\ln(\square_++\bar\Phi\Phi)+c.c.
\label{358}
\ee
Hence, we can immediately write down the answer for the effective
action (\ref{358}) by making the R-invariant shift
$G^2\to G^2+\bar\Phi\Phi$ in the action (\ref{Gamma-fin}),
\be
\Gamma_{\cN=4}=-\frac1{16\pi}\int d^3x d^2\theta
\int_0^\infty \frac{ds}{\sqrt{i\pi
s}}W^2e^{is(G^2+\bar\Phi\Phi)}\frac{\tanh(sB/2)}{sB/2}+c.c.\,,
\label{N4Gamma-fin}
\ee
or, in full $\cN=2$ superspace it reads
\bea
\Gamma_{\cN=4}&=&\frac1{2\pi}\int d^3x d^4\theta
\bigg[
-\sqrt{G^2+\bar\Phi\Phi}
+G\ln(G+\sqrt{G^2+\bar\Phi\Phi})\nn\\&&+
\frac1{4}\int_0^\infty\frac{ds}{\sqrt{i\pi s}}e^{is(G^2+\bar\Phi\Phi)}
\frac{W^2\bar W^2}{B^2}\left(
\frac{\tanh(sB/2)}{sB/2}-1
\right)\bigg].
\label{N4Gamma-fin1}
\eea
We point out that there is no Chern-Simons term induced by the
quantum corrections from $\cN=4$ hypermultiplet as soon as this
model has no parity anomaly, see, e.g., \cite{Dunne} for a review.
This was also checked in \cite{BILPSZ2} by explicit quantum
computations in $\cN=3$, $d=3$ harmonic superspace.

Similarly as in the $\cN=2$ case, the effective action
(\ref{N4Gamma-fin1}) should be superconformal. The terms in the
first line of (\ref{N4Gamma-fin1}) obviously respect the $\cN=2$
superconformal symmetry because the superfield $\bar\Phi\Phi$
transforms under the superconformal group in the same way as
$G^2$. However, the second line of (\ref{N4Gamma-fin1}) needs to
be rewritten in a superconformal form. For this purpose we
consider the following generalizations of the
superconformal quasi-primary superfields (\ref{Psin})
\bea
{\bf \Psi}&=&\frac iG\bar D^\alpha D_\alpha \ln(
G+\sqrt{G^2+\bar\Phi\Phi})\,,\nn\\
{\bf \Omega}^2&=&\frac18\frac 1{\sqrt{G^2+\bar\Phi\Phi}}\bar D^\alpha D_\alpha
\frac1G\bar D^\beta D_\beta\ln( G+\sqrt{G^2+\bar\Phi\Phi})\,.
\label{Psin4}
\eea
It is easy to see that these superfields are $\cN=2$ quasi-primary
and transform as in (\ref{delta-psi}). When the gauge multiplet is
constrained by (\ref{const-field},\ref{on-shell},\ref{backPhi}),
the superfields (\ref{Psin4}) can be represented as follows
\bea
{\bf \Psi}&=&-i\frac{\bar W^\alpha
W_\alpha}{(G^2+\bar\Phi\Phi)^{3/2}}\,,\nn\\
{\bf \Omega}^2&=&\frac18\frac{N^\alpha_\beta N_\alpha^\beta}{
(G^2+\bar\Phi\Phi)^2}
+\frac34\frac{GN^{\alpha\beta}W_\alpha\bar W_\beta}{(G^2+\bar\Phi\Phi)^3}
+\frac{15}{16}\frac{G^2W^2\bar W^2}{(G^2+\bar\Phi\Phi)^4}\,.
\eea
These representations allow us to rewrite the effective action
(\ref{N4Gamma-fin1}) in the $\cN=2$ superconformal form,
\bea
\Gamma_{\cN=4}&=&\frac1{2\pi}\int d^3x d^4\theta
\bigg[
-\sqrt{G^2+\bar\Phi\Phi}
+G\ln(G+\sqrt{G^2+\bar\Phi\Phi})\nn\\&&+
\frac1{8}\frac{{\bf \Psi}^2}{{\bf \Omega}^2}
\sqrt{G^2+\bar\Phi\Phi}
\int_0^\infty\frac{dt\,e^{it}}{\sqrt{i\pi t}}\left(
\frac{\tanh(t{\bf\Omega})}{t{\bf\Omega}}-1
\right)\bigg].
\label{N4Gamma-fin2}
\eea
Now we relax the constraint (\ref{on-shell}) and conclude that
(\ref{N4Gamma-fin2}) is the off-shell low-energy effective action
for the $\cN=4$ gauge multiplet in the long-wave approximation.

Let us single out the terms in the first line in
(\ref{N4Gamma-fin2}),
\be
\int d^3x d^4\theta
[-\sqrt{G^2+\bar\Phi\Phi}
+G\ln(G+\sqrt{G^2+\bar\Phi\Phi})]\,.
\label{dualGW}
\ee
It is easy to see that (\ref{dualGW}) can be obtained by the dimensional
reduction from the action of
$\cN=2$, $d=4$ improved tensor multiplet formulated in the
$\cN=1$, $d=4$ superspace in \cite{Rocek} which was recently
revisited in \cite{KLU}. In the three-dimensional case the action
of the form (\ref{dualGW}) was studied in \cite{HKLR}.%
\footnote{We are grateful to S.M. Kuzenko for drawing our
attention to the papers \cite{HKLR,Rocek,KLU}.}

It is interesting to note that (\ref{dualGW}) was recently obtained
 in \cite{KLL} as a dual
representation of the classical action of the Abelian Gaiotto-Witten model.
The Gaiotto-Witten model \cite{GW} is the $\cN=4$ supersymmetric
Chern-Simons-matter theory with one hypermultiplet in the
bifundamental representation of the twisted gauge group ${\cal G}_1\times
{\cal G}_2$ where the
gauge superfields corresponding to these two groups ${\cal G}_1$
and ${\cal G}_2$ have Chern-Simons rather SYM kinetic terms. The
authors of \cite{KLL} showed that in the Abelian case one of these
gauge superfields together with the hypermultiplet can be
eliminated from the classical action resulting in the action for
the second gauge superfield which appeared to have the form
(\ref{dualGW}). Hence, the classical action of the Abelian Gaiotto-Witten
model in the representation (\ref{dualGW}) arises as the leading
term in the effective action in the charged hypermultiplet model.

Our final comment is that the effective action
(\ref{dualGW}) hints the form of the effective K\"ahler
superpotential. Indeed, for the vanishing gauge superfield, $G=0$, the
expression (\ref{dualGW}) reduces to
\be
\int d^3x
d^4\theta
\sqrt{\bar\Phi\Phi}\propto
\int d^3x \frac1{\sqrt{\bar\varphi\varphi}}
\partial^m\bar\varphi\partial_m\varphi+\ldots\,,
\label{kahler} \ee where $\varphi$ is the lowest component of $\Phi$
and dots stand for the terms involving other component fields. It
would be interesting to do an independent computation of the
K\"ahler superpotential as a part of the effective action in the
three-dimensional $\cN=2$ Wess-Zumino model.

\section{Summary and discussion}
In this paper we studied the one-loop effective
action for three-dimensional $\cN=2$ and $\cN=4$ gauge superfields induced by quantum
supersymmetric matter fields. We restrict ourself to
the long-wave approximation when the
background gauge superfield is constant with respect to the
space-time coordinates and obeys the free supersymmetric Maxwell
equations. In the non-supersymmetric case such an action is
known as the Euler-Heisenberg effective action which was
studied for the three-dimensional electrodynamics in
\cite{Redlich}. The present work is a
supersymmetric generalization of the results of \cite{Redlich}.

Before computing the effective action in the model of the $\cN=2$ chiral
superfield interacting with the background gauge superfield we
found a general form (\ref{Gamma-full}) of such an action subject to the constraints
of the gauge and superconformal invariance. The leading terms in
this action are given by the Chern-Simons term (\ref{GCS}) and by
a superconformal generalization of the Maxwell action (\ref{GM}).
The functional form of these two terms is fixed by the superconformal
invariance uniquely, up to the coefficients. The higher order
terms with respect to the Maxwell field strength are taken into
account by the action (\ref{GH}) which is found up to one
arbitrary function ${\cal H}$ of quasi-primary superfields (\ref{Psin}) in the
$\cN=2$ superspace which are constructed in terms of the superfield strength
$G$ and its covariant spinor derivatives. This analysis is quite
similar to \cite{BKT} where the low-energy effective action
in the $\cN=2$, $d=4$ supergauge theory was expressed in terms of
superconformal invariants.

After considering the general structure of the superconformal action
we explicitly compute it by integrating out the chiral
superfields interacting with the background gauge superfield.
The results of the calculations match the previously proposed form
(\ref{Gamma-full}): The coefficients in Chern-Simons and Maxwell
terms are fixed as in (\ref{t1}) and (\ref{t2}) while the
higher-order contributions with respect to the Maxwell field
strengths are represented by the action (\ref{t3}) which is
expressed in terms of the quasi-primary superfields (\ref{Psin}).

The effective action for the $\cN=4$ gauge superfield
is obtained in the form (\ref{N4Gamma-fin1}).
It has no Chern-Simons term since there is no parity anomaly for
the model of charged hypermultiplet (see, e.g., \cite{Dunne} for a review).
The absence of the
Chern-Simons term in the charged hypermultiplet was also checked
in our recent work
\cite{BILPSZ2} using direct quantum computations in the $\cN=3$, $d=3$
harmonic superspace. Therefore the effective action
(\ref{N4Gamma-fin1}) starts from the Maxwell term (written in the $\cN=2$
superspace in a superconformal form) as well as contains all higher
orders of the Maxwell field strength in components. It is
interesting to note that the leading terms without derivatives in the
effective action for the $\cN=4$ gauge superfield coincide with
the classical action of the Abelian Gaiotto-Witten model rewritten in \cite{KLL}
in terms of dynamical gauge superfield. Therefore one can consider
the Abelian Gaiotto-Witten model as the effective theory
induced by quantum hypermultiplet superfield.

One of the applications of the obtained effective actions for the $\cN=2$
and $\cN=4$ gauge theories may be given within the study of the
mirror symmetry \cite{IS} for three-dimensional gauge theories. The mirror
symmetry is a kind of dualities for three-dimensional gauge
theories which relates one field theory at strong coupling with
another theory in the perturbative regime. In particular, the
leading term (\ref{t2}) in the $\cN=2$ gauge superfield effective
action is known to be dual to the K\"ahler sigma model which was
studied in \cite{BHO}. As soon as we derived not only the leading
term (\ref{t2}), but also a number of derivative contributions (\ref{t3}) in the
$\cN=2$ effective action, it is natural to find the corrections
to the sigma model considered in \cite{BHO} due to the terms
(\ref{t3}). In a similar way it would be interesting to explore the
duality for the $\cN=4$ gauge superfield effective action
(\ref{N4Gamma-fin2}). Note that modern applications of the mirror symmetry
for three-dimensional models with $\cN=2$ and $\cN=4$
supersymmetry are helpful for the studies of the ABJM-like theories \cite{JY}.

It is natural to consider the $\cN=2$ chiral superfield
interacting with the background gauge superfield and the $\cN=4$
charged hypermultiplet as the parts of the $\cN=2$
and $\cN=4$ supersymmetric three-dimensional electrodynamics, respectively. In
this case the one-loop Euler-Heisenberg-type effective actions obtained in the
present paper receive two-loop (as well as all higher-loop)
corrections which are tempting to study.
For the four-dimensional supersymmetric electrodynamics
the two-loop corrections to the supersymmetric Euler-Heisenberg effective action
were computed in \cite{Kuz03,Kuz07}, but in the three-dimensional case
this problem has never been addressed.
Finally, it is interesting to study the effective
action in the non-Abelian $\cN=2$ and $\cN=4$ three-dimensional
supergauge models and then to extend these results to the theories
with $\cN=6$ and $\cN=8$ supersymmetry which are worldvolume field
theories of M2 and D2 branes. It would open the possibility to study
the effective actions in the BLG and ABJM theories which would
give an effective quantum description of multiple M2 branes. There
are also various deformations of the BLG and ABJM models
\cite{Deform,ASW} which are interesting from the point of view of
the AdS$_4$/CFT$_3$ correspondence because they correspond to the
infrared stable superconformal points in the three-dimensional $\cN=2$
supergauge theories \cite{ASW,Penati,Kazakov}. It is natural
to study the problem of effective action in these models as well.

\vspace{3mm}
{\bf Acknowledgments}\\[3mm]
We are grateful to S.M. Kuzenko for useful comments.
I.B.S.~is indebted to O. Lechtenfeld for helpful discussions
and to ITP, Leibniz Universit\"at Hannover for kind hospitality where a
part of this work was done. The present work is supported
by RFBR grant, project No 09-02-00078 and by a
grant for LRSS, project No 3558.2010.2.
I.L.B. and I.B.S. acknowledge the support from the
RFBR grants No 10-02-90446 and No 09-02-91349 as well as from
a DFG grant, project No 436 RUS/113/669.
The work of I.B.S. is also supported by the fellowship of the Dynasty foundation.
N.G.P.\ acknowledges the support from RFBR grant, project No 08-02-00334.

\section*{Appendices}
\subsection*{A. $\cN{=}2$ superspace conventions}
\setcounter{equation}{0}
\renewcommand{\theequation}{A.\arabic{equation}}
In the present paper we use the conventions for the
three-dimensional gamma matrices following our previous works
\cite{BILPSZ2,BILPSZ1}. In particular, the gamma matrices
$(\gamma^0)_\alpha^\beta=-i\sigma_2$,
$(\gamma^1)_\alpha^\beta=\sigma_3$,
$(\gamma^2)_\alpha^\beta=\sigma_1$ obey the Clifford algebra
\be
\{ \gamma^m,\gamma^n\}=-2\eta^{mn}\,,\qquad
\eta^{mn}=\mbox{diag}(1,-1,-1)\,,
\ee
and the following orthogonality and completeness relations
\be
(\gamma^m)_{\alpha\beta}(\gamma^n)^{\alpha\beta}=2\eta^{mn}\,,\qquad
(\gamma^m)_{\alpha\beta}(\gamma_m)^{\rho\sigma}
=(\delta_\alpha^\rho\delta_\beta^\sigma+\delta_\alpha^\sigma\delta_\beta^\rho)\,.
\ee
We raise and lower the spinor indices with the $\varepsilon$-tensor, e.g.,
$(\gamma_m)_{\alpha\beta}=\varepsilon_{\alpha\sigma}(\gamma_m)^\sigma_\beta$,
$\varepsilon_{12}=1$. Any vector index can be converted into a
pair of spinor ones by the following rules
\bea
&&x^{\alpha\beta}=(\gamma_m)^{\alpha\beta} x^m\,,\qquad
x^m=\frac12(\gamma^m)_{\alpha\beta}x^{\alpha\beta}\,,\nn\\
&&\partial_{\alpha\beta}=(\gamma^m)_{\alpha\beta}\partial_m\,,\qquad
\partial_m=\frac12(\gamma_m)^{\alpha\beta}\partial_{\alpha\beta}\,,
\eea
so that
\be
\partial_m x^n=\delta_m^n\,,\qquad
\partial_{\alpha\beta} x^{\rho\sigma}
=
\delta_\alpha^\rho\delta_\beta^\sigma+\delta_\alpha^\sigma\delta_\beta^\rho
=2\delta_\alpha^{(\rho}  \delta_\beta^{\sigma)}\,.
\ee

The $\cN=2$, $d=3$ superspace is parametrized by the coordinates
$z^M=(x^m,\theta_\alpha,\bar\theta_\alpha)$ with
$\bar\theta_\alpha=(\theta_\alpha)^*$. The covariant spinor
derivatives
\be
D_\alpha=\frac\partial{\partial\theta^\alpha}+i\bar\theta^\beta
\partial_{\alpha\beta},\qquad
\bar D_\alpha=-\frac\partial{\partial\bar\theta^\alpha}
-i\theta^\beta \partial_{\alpha\beta}
\ee
obey the standard anticommutation relation
\be
\{D_\alpha, \bar D_\beta
\}=-2i\partial_{\alpha\beta}\,.
\label{Dalg}
\ee
The integration measure in the full $\cN=2$, $d=3$ superspace is
defined as
\be
d^7z\equiv d^3x d^4\theta=\frac1{16}d^3x\,D^2\bar D^2\,,\quad
\mbox{so that}\quad
\int d^3x\, f(x)=\int d^7z\,\theta^2\bar\theta^2 f(x)\,,
\label{fullmeasure}
\ee
for some field $f(x)$. Here we use the following conventions for
contractions of the spinor indices
\be
D^2= D^\alpha D_\alpha\,,\quad
\bar D^2=\bar D^\alpha\bar D_\alpha\,,\quad
\theta^2=\theta^\alpha\theta_\alpha\,,\quad
\bar\theta^2=\bar\theta^\alpha\bar\theta_\alpha\,.
\ee

The chiral subspace is parametrized by $z_+=(x_+^m,\theta_\alpha)$,
where $x_\pm^m=x^m\pm i\gamma^m_{\alpha\beta}\theta^\alpha\bar\theta^\beta$.
The chiral superfields are defined as usual, $\bar
D_\alpha\Phi=0$ $\Rightarrow$ $\Phi=\Phi(x_+^m,\theta_\alpha)$. The
integration measure in the chiral superspace
$d^5z\equiv d^3x d^2\theta$ is related to the
full superspace measure (\ref{fullmeasure}) as
\be
d^7z=-\frac14d^5z\,\bar D^2\,.
\label{chiral-measure}
\ee

\subsection*{B. Superconformal transformations in $\cN=2$ superspace}
\setcounter{equation}{0}
\renewcommand{\theequation}{B.\arabic{equation}}
Here we review a representation of the superconformal group on the
superfields in the $\cN=2$, $d=3$ superspace which was used in Section 2.2
(see some details in \cite{Park}, the analogous construction
for $\cN=1$, $d=4$ superspace was given in \cite{bookBK}).

Let us consider the infinitesimal superconformal transformations of coordinates
of the $\cN=2$ superspace
$z^A=(x^{\alpha\beta},\theta^\alpha,\bar\theta^\alpha)$,
\be
z^A \longrightarrow z^A+\delta_{\rm sc} z^A\,,
\label{change}
\ee
where $\delta_{\rm sc} z^A$ explicitly reads
\bea
\delta_{\rm sc} x^{\alpha\beta}&=&ax^{\alpha\beta}
+x^{\alpha\rho}x^{\beta\gamma}k_{\rho\gamma}
-\frac12\theta^2\bar\theta^2 k^{\alpha\beta}
+2i\theta^{(\alpha}x^{\beta)\gamma}\eta_\gamma
+\theta^2\bar\theta^{(\alpha}\eta^{\beta)}
\nn\\&&
+2i\bar\theta^{(\alpha}x^{\beta)\gamma}\bar\eta_\gamma
+\bar\theta^2\theta^{(\alpha}\bar\eta^{\beta)}\,,
\label{delta-x}
\\
\delta_{\rm sc}\theta^\alpha&=&(a/2+ib)\theta^\alpha
 +\theta^\beta x^{\alpha\gamma}k_{\beta\gamma}
 +\frac i2\theta^2\bar\theta^\beta k^\alpha_\beta
 +i\theta^2\eta^\alpha+(x^{\alpha\beta}
 +2i\theta^{(\alpha}\bar\theta^{\beta)})\bar\eta_\beta\,,
\label{delta-theta}\\
\delta_{\rm sc}\bar\theta^\alpha&=&(a/2-ib)\bar\theta^\alpha
 +\bar\theta^\beta x^{\alpha\gamma} k_{\beta\gamma}
 +\frac i2\bar\theta^2\theta^\beta k^\alpha_\beta
 +i\bar\theta^2\bar\eta^\alpha
 +(x^{\alpha\beta}-2i\theta^{(\alpha}\bar\theta^{\beta)})\eta_\beta\,.
\label{delta-bartheta}
\eea
Here $a$, $b$, $k_{\alpha\beta}$, $\eta_\alpha$,
$\bar\eta_\alpha$ are the parameters of dilatations,
U(1) transformations, special conformal transformations and
S-supersymmetry transformations, respectively. These
transformations can be shown to obey the superconformal algebra
$osp(2,\mathbb{R}|2)$.

Let us consider a superconformal
Killing vector $\xi^A=\xi^A(z)=(\xi^{\alpha\beta}(z),
\xi^\alpha(z),\bar\xi^\alpha(z))$, where
\be
\xi^{\alpha\beta}=\delta_{\rm sc} x^{\alpha\beta}
-2i\delta_{\rm sc}\theta^{(\alpha}\bar\theta^{\beta)}
+2i\theta^{(\alpha}\delta_{\rm sc}\bar \theta^{\beta)}\,,
\quad
\xi^\alpha=\delta_{\rm sc}\theta^\alpha\,,
\quad
\bar\xi^\alpha=\delta_{\rm sc}\bar\theta^\alpha\,.
\ee
The explicit expressions for the components of the superconformal
Killing vector can be derived from
(\ref{delta-x})--(\ref{delta-bartheta}),
\begin{eqnarray}
\xi^{\alpha\beta}&=&ax^{\alpha\beta}
+4b\theta^{(\alpha}\bar\theta^{\beta)}
+ k_{\gamma\delta}x_+^{\gamma(\alpha}x_-^{\beta)\delta}
-2ix^{\gamma(\alpha}k^{\beta)}_\gamma\theta^\rho\bar\theta_\rho
\nn\\&&
 +4i\theta^{(\alpha}x_-^{\beta)\gamma}\eta_\gamma
 +4i\bar\theta^{(\alpha}x_+^{\beta)\gamma}\bar\eta_\gamma
 \,, \label{zeta-ab}\\
\xi^\alpha&=&(a/2+ib)\theta^\alpha
+ k_{\gamma\delta} x_+^{\alpha\gamma}\theta^\delta
+i\theta^2\eta^\alpha
+x_+^{\alpha\beta}\bar\eta_\beta\,, \label{zeta} \\
\bar\xi^{\alpha}&=&(a/2-ib)\bar\theta^{\alpha}
 + k_{\gamma\delta}x_-^{\alpha\gamma}\bar\theta^\delta
+i\bar\theta^2\bar\eta^\alpha
+x_-^{\alpha\beta}\eta_\beta\,,\label{bar-zeta}
\end{eqnarray}
where
\be
x_\pm^{\alpha\beta}=x^{\alpha\beta}\pm2i\theta^{(\alpha}\bar\theta^{\beta)}\,.
\ee
There is a superform $\xi$ associated with the superconformal
Killing vector,
\be
\label{xi}
\xi=
\xi^A D_A=\frac12\xi^{\alpha\beta}(z)\partial_{\alpha\beta}
+\xi^\alpha(z) D_\alpha-\bar\xi^\alpha(z) \bar D_\alpha\,,
\ee
which obeys the following important relation
\be
[\xi,D_\alpha]\propto D_\beta\,.
\label{propo}
\ee
Equation (\ref{propo}) means that the superconformal
transformations respect the chirality and hence they can be
extended to chiral superfields.

Either from (\ref{propo}) or from the explicit relations  (\ref{zeta-ab})--(\ref{bar-zeta})
one can deduce the following properties for spinor components of
the superconformal Killing vector,
\bea
&&\xi^\alpha=\frac i6\bar D_\beta \xi^{\alpha\beta}\,,\qquad
\bar D_\alpha\xi_\beta=0\,,\nn\\
&&\bar \xi^\alpha=-\frac i6 D_\beta \xi^{\alpha\beta}\,,
\qquad D_\alpha\bar \xi_\beta =0\,,
\label{xi-chiral}
\eea
and for the vector one,
\be
D^2\xi^{\alpha\beta}=\bar D^2\xi^{\alpha\beta}=0\,,\qquad
D^{(\alpha}\xi^{\beta\gamma)}=\bar
D^{(\alpha}\xi^{\beta\gamma)}=0\,.
\label{208}
\ee
The relations (\ref{xi-chiral}) show that $\xi^\alpha$ and
$\bar\xi^\alpha$ are chiral and antichiral, respectively. Moreover,
they are expressed in terms of the vector component
$\xi^{\alpha\beta}$ which satisfies (\ref{208}). In fact, the
equations (\ref{208}) are main defining relations for the
superconformal Killing vector which lead to the standard equation
for $\xi^m=\frac12\gamma^m_{\alpha\beta}\xi^{\alpha\beta}$,
\be
\partial_m \xi_n+\partial_n\xi_m =\frac23\eta_{mn}\partial_p
\xi^p\,.
\ee
As is shown in \cite{Park}, one can in principle start with
(\ref{propo}), then deduce (\ref{xi-chiral}) and (\ref{208}) and
after that derive (\ref{zeta-ab}) as a solution of (\ref{208}).%
\footnote{Note that (\ref{zeta-ab}) is not the general solution of (\ref{208}).
The general solution involves also the parameters of usual translations,
Lorentz transformations and supertranslations. However we omit here these
parameters since the superspace approach provides the covariance of
all considered actions under super Poincar\'e group.}

We will need also the following properties of the
components of the superconformal Killing vector
\bea
D^{(\alpha}\xi^{\beta)}+\bar D^{(\alpha}\bar
\xi^{\beta)}&=&0\,,
\label{206}\\
D^\alpha\xi_\alpha-\bar D^\alpha\bar\xi_\alpha&=&-\frac13\partial_{\alpha\beta}
\xi^{\alpha\beta}=-2\rho\,,
\label{207}
\eea
where
\be
\rho=a+k_{\alpha\beta}x^{\alpha\beta}
 +2i\theta^\alpha\eta_\alpha
 +2i\bar\theta^\alpha\bar\eta_\alpha\,.
\label{sigma}
\ee
The superfield $\rho$ obeys
\be
D_\alpha\rho
 =2i(k_{\alpha\beta}\bar\theta^\beta+\eta_\alpha)\,,\qquad
\bar D_\alpha\rho
 =-2i(k_{\alpha\beta}\theta^\beta+\bar\eta_\alpha)\,,
\label{sig1}
\ee
while the following second derivatives of $\rho$ vanish
\be
D^2\rho= \bar D^2\rho =D^\alpha\bar D_\alpha\rho =0\,.
\label{sig2}
\ee

Let us introduce also the expressions
\bea
\sigma&=&\frac14\partial_{\alpha\beta}\xi^{\alpha\beta}
-\frac12D_\alpha \xi^\alpha
=a-ib+k_{\alpha\beta}x_+^{\alpha\beta}+4i\theta^\alpha\eta_\alpha\,,
\label{rho}
\\
\bar\sigma&=&\frac14\partial_{\alpha\beta}\xi^{\alpha\beta}
+\frac12\bar D_\alpha \bar\xi^\alpha
=a+ib+k_{\alpha\beta}x_-^{\alpha\beta}+4i\bar\theta^\alpha\bar\eta_\alpha\,.
\label{bar-rho}
\eea
These superfields are chiral and antichiral, respectively,
\be
\bar D_\alpha \sigma=0\,,\qquad
D_\alpha \bar \sigma=0\,.
\label{rho-chiral}
\ee
Clearly, the parameters $\sigma$, $\bar\sigma$ and $\rho$ are
related to each other,
\be
\label{rho-sigma}
\rho=\frac12(\sigma+\bar\sigma)\,.
\ee

Now we consider a representation of the superconformal group on
superfields in the $\cN=2$ superspace. Given a (real) superfield $V$
defined in full $\cN=2$ superspace with mass-dimension $l$,
we define its infinitesimal superconformal transformation as
\be
\delta_{\rm sc} V=(l\rho+\xi)V\,,
\label{dV}
\ee
where $\xi$ and $\rho$ are defined in (\ref{xi}) and
(\ref{sigma}), respectively. Analogously, the superfields
$\sigma$ and $\bar\sigma$ given by (\ref{rho},\ref{bar-rho}) are
used to define the superconformal transformations for a chiral
$Q$ and an antichiral $\bar Q$ superfields of mass dimension $l$,
\be
\delta_{\rm sc} Q =(l\sigma+\xi)Q\,,\qquad
\delta_{\rm sc} \bar Q=(l\bar\sigma+\xi)\bar Q\,.
\label{dQ}
\ee
The superfields which transform under the superconformal group by
the rules (\ref{dV},\ref{dQ}) are usually referred to as the
quasi-primary superfields with scaling dimension $l$.

An action $S=\int d^3x d^4\theta\,{\cal L}$ is superconformal
if the Lagrangian $\cal L$ transforms as a quasi-primary
scalar superfield with the scaling dimension $l=+1$, i.e.,
\be
\delta_{\rm sc} {\cal L}=(\rho+\xi){\cal L}
\quad\Rightarrow\quad
\delta_{\rm sc} S=0
\,.
\label{inv-act}
\ee
However, this is not the necessary and sufficient condition of
superconformal invariance since more generally the Lagrangian can transform
as
\be
\delta_{\rm sc} {\cal L}=(\rho+\xi){\cal L}
+\sigma {\cal K}+\bar\sigma \bar{\cal K}
\quad\Leftrightarrow\quad
\delta_{\rm sc} S=0
\,,
\label{inv-act1}
\ee
where $\cal K$ and $\bar{\cal K}$ are linear functions,
\be
\bar D^2{\cal K}=0\,,\qquad
D^2\bar{\cal K}=0\,.
\ee
These additional contributions with ${\cal K}$ and $\bar{\cal K}$ in the
variation of the Lagrangian (\ref{inv-act1}) do not break the invariance of the
action since they vanish in passing from the full to the (anti)chiral
superspace.

Similarly, the
superconformal invariance of an action in the chiral superspace
$S_c=\int d^3x d^2\theta\,{\cal L}_c$ is guaranteed if the
Lagrangian ${\cal L}_c$ is a chiral quasi-primary superfield with the scaling dimension $l=+2$,
\be
\delta_{\rm sc} {\cal L}_c=(2\sigma+\xi){\cal L}_c
\quad\Leftrightarrow\quad
\delta_{\rm sc} S_c=0\,.
\ee

\end{document}